\documentclass[journal]{IEEEtran}

\ifCLASSINFOpdf
   \usepackage[pdftex]{graphicx}
\else
\fi

\usepackage{xspace}
\usepackage{booktabs} 
\usepackage{xcolor}
\usepackage{color}
\usepackage{colortbl}
\usepackage{pifont}
\definecolor{darkgreen}{rgb}{0.0, 0.5, 0.0}
\definecolor{babyblueeyes}{rgb}{0.63, 0.79, 0.95}
\newcommand{\cmark}{\textcolor{darkgreen}{\ding{51}}}

\newcommand{\xmark}{\textcolor{red}{\ding{55}}}

\usepackage{multirow}
\usepackage{paralist}

\def\subparagraph{} 
\usepackage[compact]{titlesec}

\usepackage{listings}
\newcommand{\noind}[0]{\par \noindent}
\newcommand{\noindpar}[1]{\noind {\bf #1}}

\usepackage[scale=.9,light]{sourcecodepro}
\usepackage[T1]{fontenc}
\newcommand{\code}[1]{{\ttfamily #1}}

\newcommand{\vmname}{PureVM\xspace}
\newcommand{\langname}{PureLANG\xspace}
\newcommand{\implI}{RewindingVM\xspace}
\newcommand{\implII}{JustInTimeVM\xspace} 
\newcommand{\implIII}{TestVM\xspace}

\usepackage{comment}
\usepackage{lipsum}
\usepackage{paralist}

\begin{document}

\title{Virtualizing Intermittent Computing}

\author{\c{C}a\u{g}lar~Durmaz,~Kas{\i}m~Sinan~Y{\i}ld{\i}r{\i}m,~\IEEEmembership{IEEE Member,}~and~Geylani~Kardas
\thanks{\c{C}a\u{g}lar Durmaz and Geylani Kardas are with the International Computer Institue, Ege University, Turkey. e-mail: \{caglar.durmaz, geylani.kardas\}@ege.edu.tr. Kas{\i}m Sinan Y{\i}ld{\i}r{\i}m is with the Department of Information Engineering and Computer Science, University of Trento, Italy. e-mail: kasimsinan.yildirim@unitn.it}.
}


\maketitle

\begin{abstract}

Intermittent computing requires custom programming models to ensure the correct execution of applications despite power failures. However, existing programming models lead to programs that are hardware-dependent and not reusable. This paper aims at virtualizing intermittent computing to remedy these problems. We introduce \vmname, a virtual machine that abstracts a transiently powered computer, and \langname, a continuation-passing-style programming language to develop programs that run on \vmname. This virtualization, for the first time, paves the way for portable and reusable transiently-powered applications.

\end{abstract}

\begin{IEEEkeywords} 
Energy Harvesting, Batteryless IoT,
Intermittent Computing, Domain-specific Language, Virtual Machine
\end{IEEEkeywords}

\IEEEpeerreviewmaketitle

\section{Introduction}

\IEEEPARstart{I}{n} the past decade, the progress in energy harvesting circuits and the decrease in power requirements of processing, sensing, and communication hardware promised the potential of freeing the Internet of Things (IoT) devices from their batteries. Recent works demonstrated several microcontroller-based devices that can work without the need for batteries by harvesting energy from ambient sources, such as solar and radiofrequency~\cite{hester2017flicker,yildirim2018safe,nardello2019camaroptera}. Batteryless devices store the harvested ambient energy into a tiny capacitor that powers the microcontroller and peripherals. A batteryless device can compute, sense, and communicate when the energy stored in its capacitor is above an operating threshold. It turns off and loses its volatile state (e.g., the contents of the CPU, peripheral registers, the volatile memory) when the energy level drops below this threshold. The device can turn on only after charging its capacitor again. This phenomenon, i.e., the intermittent execution due to power failures, led to the emergence of a new computing paradigm, the so-called \emph{intermittent computing}~\cite{yildirim2018ink,kortbeek2020TICS}. 

During intermittent execution, batteryless devices use the harvested energy to perform a short burst of computation.  To recover their computation state and progress computation forward after a power failure, they need to save the computation state in non-volatile memory before a power failure. Recent studies proposed programming models for intermittent computing to support these state logging and recovery operations. The proposed programming models provide language constructs (i.e., either \emph{checkpoints}~\cite{kortbeek2020TICS} or \emph{tasks}~\cite{colin2016chain}) to (i) maintain the forward progress of computation and (ii) keep the memory (i.e., computation state) consistent. However, with these models, programmers need to deal with the low-level details of the intermittent execution~\cite{durmaz2019puremem}. In particular, existing models pose the following deficiencies.

\noindpar{Explicit Burst Management.} Programmers need to design their programs as a set of computation bursts that should fit in the capacitor. Thus, they explicitly identify the boundaries of these bursts via checkpoint placement or task decomposition. This situation increases the programming effort considerably.

\noindpar{Hardware Dependency.} The active time of a batteryless device depends on its capacitor size and its power consumption, i.e., its hardware configuration~\cite{colin2018reconfigurable}. Programmers might need to identify different burst boundaries to execute their programs on a new device with a different hardware configuration. Therefore, existing intermittent programs are not portable, and in turn, not reusable. 
    
\noindpar{Explicit I/O Management.} Power failures that occur during I/O or interrupt handling might leave the memory in an inconsistent state. With existing models, programmers need to manually ensure the atomic execution of interrupt handlers or I/O operations~\cite{ruppel2019Coati}. This situation increases the programming burden and makes programs error-prone. 

\begin{figure}
    \centering
    \includegraphics[width=1\columnwidth]{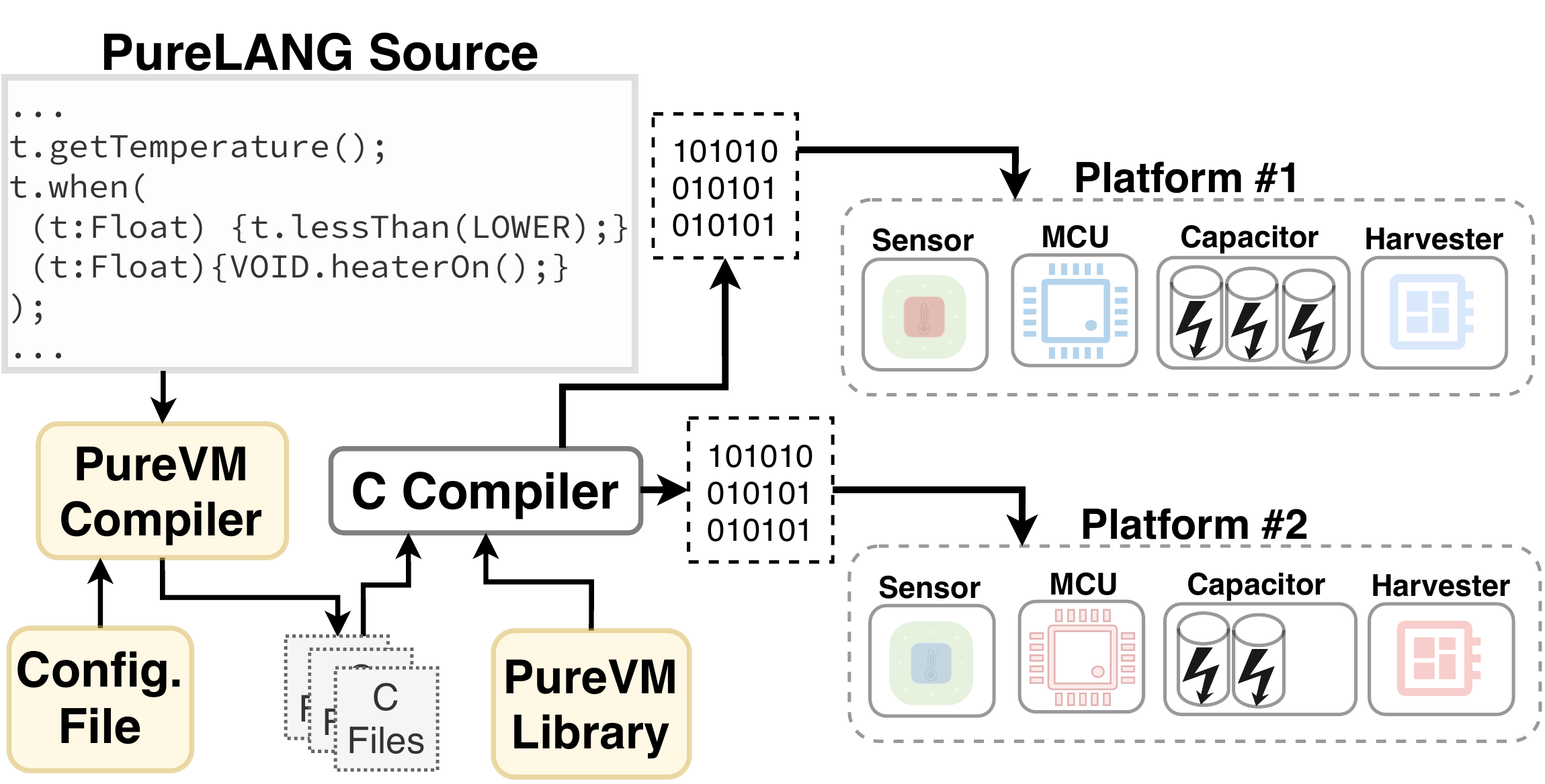}
    \caption{Using \langname, a continuation-passing-style programming language, and \vmname, whose specification enables the execution of programs written in \langname, programmers do not reason about the intermittent execution and they develop programs portable across several hardware platforms.}
    \label{fig:intro}
\end{figure}

In this paper, we aim at virtualizing intermittent computing to remedy the deficiencies mentioned above. We introduce \vmname virtual machine and its software interface \langname language that abstract away the complicated aspects of intermittent execution. Thanks to \vmname and \langname, programmers focus only on their application logic, forget about power failures as if they are programming continuously powered systems, and develop portable intermittent programs. Shortly, this paper introduces the following contributions:  
\begin{compactenum}
    \item \textbf{\vmname Virtual Machine.} We introduce \vmname, the first \emph{virtual machine} for intermittent systems, which abstracts a transiently powered computer. This abstraction hides the details of intermittent execution and enables platform-independent program development via its implementations targeting different hardware (see Figure~\ref{fig:intro}). 
    
    
    \item \textbf{\langname Language.} We introduce \langname, a  \emph{continuation-passing-style} programming language, intended to prevent programmers from reasoning about intermittent execution.  
    \langname programs are translated into a set of re-composable and atomically-executed \vmname instructions.  
    
\end{compactenum}
To the best of our knowledge, our work proposes the first \emph{virtualization} attempt for intermittent computing. \vmname and \langname create the fundamental building blocks of intermittent computing from scratch that pave the way for portable transiently-powered applications.

\section{Background and Related Work}
\label{sec:background}

Frequent power failures lead to intermittent execution that poses several challenges to developing batteryless applications. We classify these challenges into two classes: computation and I/O related challenges (A1--A3) and programming challenges (B1--B2). We describe them as follows.

\noindpar{A1- Non-termination and Memory Consistency.} Power failures hinder the \emph{forward progress} of the computation, which leads to non-terminating programs~\cite{maeng2018Chinchilla}. Non-termination occurs since the device loses the intermediate results and restarts the computation from scratch upon each reboot. Power failures might also lead to \emph{memory inconsistencies}~\cite{ransford2014nonvolatile}. To give an example, assume that a program is modifying the persistent variable \code{var} (i.e., a variable kept in non-volatile memory) by executing the code block \code{\{vector[var++]=10;\}}. Since the variable \code{var} is updated after it is read, there is a \emph{Write-After-Read  (W.A.R.)} dependency.  If a power failure occurs at this point, the re-execution of this code will increment the variable \code{var} again. Therefore, another element of the \code{vector} will be set to 10 in this case. Due to the W.A.R. dependency, there is a violation of \emph{idempotency} since repeated computation produces different results. To prevent these issues, programmers should divide their programs into a set of idempotent code blocks (e.g., \emph{tasks}~\cite{colin2016chain}) that can execute \emph{atomically} in a power cycle and can be safely restarted despite W.A.R. dependencies. A runtime library (e.g., \cite{alpaca,yildirim2018ink}) is required to persist the program state non-volatile memory, to manage power failures, and to re-execute the code blocks that could not complete in the previous power cycle.

\noindpar{A2- Control-flow Inconsistencies.}  If the control flow depends on external inputs such as sensor readings, power failures might lead to erratic program behavior~\cite{surbatovich2019dependent}. In particular, programmers need to pay special attention to implementing conditional statements that check persistent variables, whose values might be updated during I/O operations. For example, consider the case that a program reads a temperature value (\code{temp = read\_sensor();}) and sets the variable \code{alarm} based on the temperature reading (\code{if(temp > limit) then alarm = true; else tempOk = true;}). If the temperature is less than a pre-defined limit, the variable \code{tempOK} will be set to true. If there is a power failure right after this operation and the program re-executes, the program might read another temperature value higher than the limit. In this case, the program will set the variable \code{alarm} to true. At this point, both of the variables \code{alarm} and \code{tempOK} are true, which is logically incorrect~\cite{surbatovich2019dependent}. 

\noindpar{A3- Handling Interrupts.} Interrupts cause dynamic branches, which move the control from the main thread of execution to the interrupt service routine. The main program and interrupt service routines might share the persistent state. If an interrupt service routine leaves the shared persistent variables partially updated due to a power failure, this situation might lead to memory inconsistencies~\cite{ruppel2019Coati}.  

\noindpar{B1- Platform Dependencies.} The execution time of an intermittent program depends on several factors, such as available harvestable energy in the environment, capacitor size (i.e., energy storage capacity), and the energy consumption profile of hardware and software. Intermittent programs need to be modified and restructured regarding these factors to eliminate non-termination and ensure computational progress~\cite{maeng2018Chinchilla}. 

\noindpar{B2- Reuse and Maintaining Difficulties.} Platform and runtime dependencies make implementing reusable intermittent programs difficult~\cite{durmaz2019puremem}. For example, programmers using task-based models need to deal with task decomposition and task-based control flow~\cite{colin2016chain}. Handling these issues is complicated and leads to programs that are difficult to maintain.

\subsection{The State of the Art}

We classify the prior art based on how they addressed the aforementioned challenges.

\noindpar{Checkpoint-based Systems.} 
Checkpointing runtime environments (e.g.,~\cite{ransford2012mementos,lucia2015DINO,jayakumar2014quickrecall,balsamo2016hibernus++}) persist the registers, global variables, and the call-stack of programs into non-volatile memory via programmer-inserted checkpoints to preserve the forward progress of computation (A1).  In particular, due to the call stack's dynamic nature (i.e., it grows and shrinks dynamically), the programmers need to place checkpoints carefully to eliminate non-termination. In particular, the energy required to execute the instructions between two checkpoints should not exceed the maximum energy stored in the capacitor. Therefore, checkpoint placement is platform-dependent and checkpointed programs are not reusable. There are studies (e.g.,~\cite{woude2016ratchet,mottola2017harvos,kortbeek2020TICS,maeng2020CatNap,maeng2019Samoyed,maeng2018Chinchilla}) that provide compilers to translate C programs into intermittent programs without programmer intervention. However, the C language does not provide abstractions for interrupt handling (A3) and atomic I/O (A2) operations on intermittent systems. The absence of these abstractions might lead to memory inconsistencies and non-termination.

\noindpar{Task-based Systems.} Task-based models (e.g.,~\cite{colin2016chain,alpaca,yildirim2018ink,hester2017mayfly,majid2020Coala,ruppel2019Coati}) require programmers to structure their programs as a collection of idempotent and atomic tasks. They eliminate the need for the call-stack and checkpoints by employing GOTO-style transitions among tasks, i.e., task-based control-flow. However, this is an unstructured programming style that leads to programs that are not reusable and that are prone to bugs~\cite{dijkstra1968letters}. Task-based programming also leads to platform-dependent code since task sizes depend on the capacitor size of the platform. Recent work~\cite{durmaz2019puremem} provides tasks with parameters and continuation-passing~\cite{sussman1998scheme} via closures, which enables reusable code by delivering the control flow in a structured way, similar to function calls. However, it also leads to platform-dependent code because of static task sizes.

\begin{table}
\caption{The state of the art runtimes and mentioned challenges.}
\label{table:runtime_comparison}
\centering
\begin{tabular}{>{\centering}m{0.21\textwidth}
			>{\centering}m{0.07\textwidth}
			>{\centering}m{0.009\textwidth}
			>{\centering}m{0.009\textwidth}
			>{\centering}m{0.009\textwidth}
			>{\centering}m{0.009\textwidth}
			m{0.01\textwidth}<{\centering}}
\toprule
  \textbf{Runtime} & \textbf{Type} & \textbf{A1} & \textbf{A2} & \textbf{A3} & \textbf{B1} & \textbf{B2}\\ 
                                     
\toprule

QuickRecall\cite{jayakumar2014quickrecall}, Mementos\cite{ransford2012mementos} 
                            & Checkpt. & \xmark & \xmark & \xmark & \xmark & \cmark\\
                \rowcolor{gray!10}
DINO\cite{lucia2015DINO}, Ratchet\cite{woude2016ratchet}, Chinchilla\cite{maeng2018Chinchilla}     
                            & Checkpt & \cmark & \xmark & \xmark & \xmark & \cmark\\
Samoyed\cite{maeng2019Samoyed}
                            & Checkpt. & \cmark & \cmark &\xmark & \xmark & \cmark\\
\rowcolor{gray!10}                              
TICS\cite{kortbeek2020TICS}, CatNap\cite{maeng2020CatNap}
                            & Checkpt. &\cmark &\xmark &\cmark &\xmark &\cmark\\
Chain\cite{colin2016chain}, Alpaca\cite{alpaca}, Mayfly\cite{hester2017mayfly}, Coala\cite{majid2020Coala} Rehash~\cite{bakar2021rehash}
                            & Task-based &\cmark &\xmark&\xmark&\xmark &\xmark\\
 \rowcolor{gray!10} 
PureMEM\cite{durmaz2019puremem}
                            & Task-based & \cmark &\xmark &\xmark &\xmark & \cmark\\
Coati\cite{ruppel2019Coati}, InK\cite{yildirim2018ink}  
                            & Task-based & \cmark &\xmark & \cmark &\xmark &\xmark\\
                \rowcolor{yellow!30} 
{{\bf This Work} (\vmname/\langname) }
                            & Virtual Machine &\cmark &\cmark &\cmark &\cmark &\cmark\\
\bottomrule
\end{tabular}

\end{table}

\subsection{Our Differences}  

Table \ref{table:runtime_comparison} provides a comparison of our work with the state-of-the-art. We propose an intermittent computing solution composed of a virtual machine (\vmname), a programming language (\langname), and a compiler. We design \vmname to abstract the intermittent execution details and give the programmer a continuously powered view of the target device. This abstraction provides platform-independent code via its multiple compilers for multiple hardware. \langname is the software interface of \vmname. \langname programs are translated into a sequence of \emph{primitive functions}, which are the smallest computation blocks that \vmname executes atomically despite power failures. Thanks to this two-layered abstraction, our work overcomes all mentioned challenges of intermittent computing (i.e., A1--A3 and B1--B2).
\section{\langname Language}
\label{sec:language}

\langname is a statically-typed and event-driven programming language. Programmers develop \langname applications via objects and functions that operate on them, and do not reason about intermittent execution. \langname employs continuation-passing style~\cite{sussman1998scheme} where the program control flow is passed explicitly via \emph{continuations}. Each function (or expression) takes one \emph{flow-in object} in addition to its parameters and passes one \emph{flow-out object} to the next function (or expression) that handles the rest of the computation. Therefore, each continuation contains a reference to a flow-in object,  a set of object references as parameters, and a function reference to be applied. Events are continuations that are created by interrupt handlers. 

\vmname persists continuations in non-volatile memory to ensure forward progress. The overhead of persisting a continuation is static (since it contains only a function reference and a certain number of object references) compared to persisting the call-stack whose size is dynamic in procedural languages. During program execution, \vmname applies the function on the flow-in object by using the information stored in the current continuation. Since \langname functions always operate on objects (kept in non-volatile memory), \vmname can track the updates on these objects to preserve data consistency despite power failures. 

\begin{figure}
	\centering
	\includegraphics[width=0.85\linewidth]{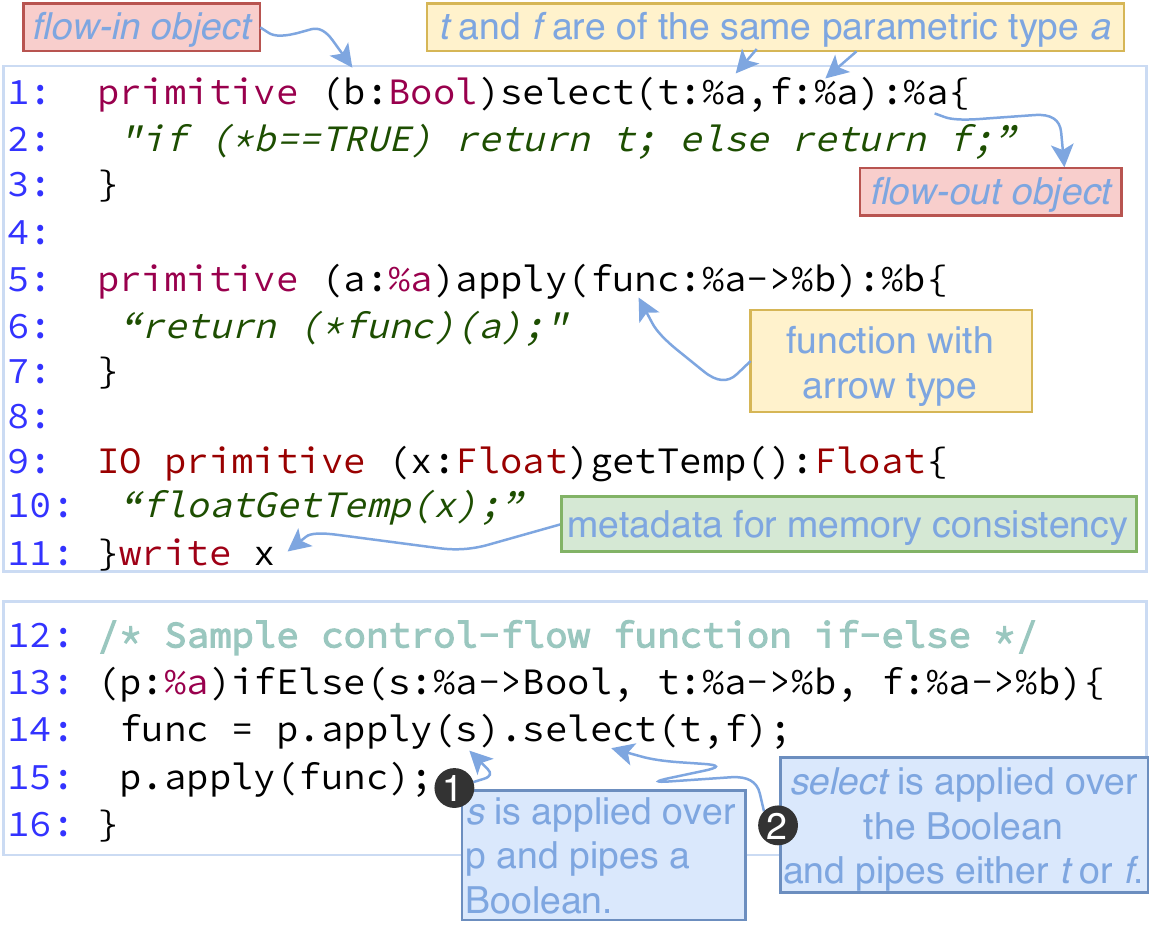}
	\caption{\langname example primitive and control-flow functions. The control-flow only depends on two primitive functions \textit{apply} and \textit{select} in \langname.}
	\label{fig:control_flow}
\end{figure}

\subsection{\langname Types and Primitive Functions}

\label{sec:io:primitive}

Primitive functions are the high-level instructions, which execute atomically and form the interface between the \langname and \vmname. As an analogy with task-based systems~\cite{colin2016chain,yildirim2018ink,alpaca,ruppel2019Coati,hester2017mayfly,bakar2021rehash}, primitive functions are the atomic tasks that are reusable building blocks. \langname programs are a composition of these atomic blocks. 

\noindpar{Parameteric and Arrow Type.} \langname has built-in object types \code{Int}, \code{Float}, \code{Bool} and \code{Void}, which are reference types that point an address in memory. Moreover, \langname offers \emph{parametric types} that are used to specify the type later. As an example, the \code{select} primitive function (Fig.~\ref{fig:control_flow}, Lines 1--3) returns one of two objects \code{t} or \code{f} of same parametric type \code{\%a} as a flow-out object. The objects referenced by \code{t} and \code{f} can be of any type but same. The primitive function \code{select} makes the decision based on the value of the boolean flow-in object \code{b}. Functions with the parametric type (also known as \emph{parametric polymorphism}) eliminate code duplication. Without parametric polymorphism, \code{select} method needs different implementations for different types. As another example, the  primitive function \code{apply} (Fig.~\ref{fig:control_flow}, Line 5) applies the given function on the flow-in object of parametric type \code{\%a} and  returns a flow-out object of parametric type \code{\%b}. It also takes a function reference \code{func} as a parameter, which takes an object of  parametric type \code{\%a} and returns an object of parametric type \code{\%b}. This is indicated by using \emph{arrow type} decleration depicted as \code{\%a->\%b}. In the body of the primitive function (Fig.~\ref{fig:control_flow}, Line 6), \code{func} is called by passing the flow-in object \code{a}. Note that, \code{func} returns an object of type \code{\%b}, which is compatible with the flow-out object type of  \code{apply}. 

\noindpar{IO Primitives.} \langname introduces IO primitive function functions to eliminate control-flow inconsistencies during I/O operations (A2). \code{IO} metadata (e.g., see \code{getTemp} function in Fig.~\ref{fig:control_flow}, Line 9) helps the \langname compiler to handle these operations differently. The compiler splits \langname code blocks with IO primitives into three sections: pre-IO primitive, IO primitive itself, and post-IO primitive. After each section executes, \vmname takes control to persist computational state, which ensures the atomic execution of the IO primitive. 

\noindpar{Type Checking.} Arrow and parametric type declarations help the \langname compiler for type inference and type checking. While decomposing the program into its primitive functions, the \langname compiler performs type checking by using input and output type metadata to eliminate (B2)-type bugs. The compiler also infers the variable types automatically when the programmer does not declare them explicitly. 

\noindpar{Resolving W.A.R. Dependencies.} Primitive functions also specify a meta-data concerning write operations on objects. As an example, the \textit{write} in the definition of \code{getTemp} function tells the compiler that this function modifies the flow-in object \code{x}. While decomposing the program into its primitive functions,  the compiler can resolve W.A.R. dependencies using this metadata. This situation helps \vmname to execute the intermittent program correctly by preserving the memory consistency (to ensure A1). Considering the target \vmname implementation, \langname compiler instruments the bodies of the functions by inserting the necessary undo logging or redo logging code explained in Section~\ref{sec:vm}.

\subsection{\langname Statements and Control Flow}

Since \langname employs structured programming, complex expressions are formed by composing other expressions (and primitive functions). The dot operator (\code{.}) enables expression composition as long as the output type of sub-expression is compatible with the input type of the following sub-expression. The last statement in a function body determines the output object that should be compatible with the output type of the function. Thanks to the continuation-passing style of the language, all statements forming the complex behavior of a function execute in order. Therefore, there is no need for the \vmname to check branches and early exits.

\noindpar{Control Flow.} In \langname,  every function related to the control flow is a composition of \code{select} and \code{apply} primitive functions. For example, \code{ifElse} function (Fig.~\ref{fig:control_flow}, Lines 14--16) enables a conditional branch by invoking the \code{apply} and \code{select} primitive functions in order. The first parameter \code{s} is a function that takes an object of parametric type \code{\%a} and returns a boolean object. Firstly, the function \code{s} is applied on the flow-in object \code{p}, which pipes a boolean object (i.e., \code{p.apply(s)} in Line 11 which returns boolean). Then, the returned object becomes a flow-in object for the \code{select} primitive, which returns one of the functions from \code{t} and \code{f} by considering the flow-in boolean object. The returned function object is assigned to variable \code{func}. Then, \code{func} is applied to the flow-in object \code{p}, and an object of \code{\%b}  type is returned (Line 12).

\begin{figure}
	\centering
	\includegraphics[width=0.95\linewidth]{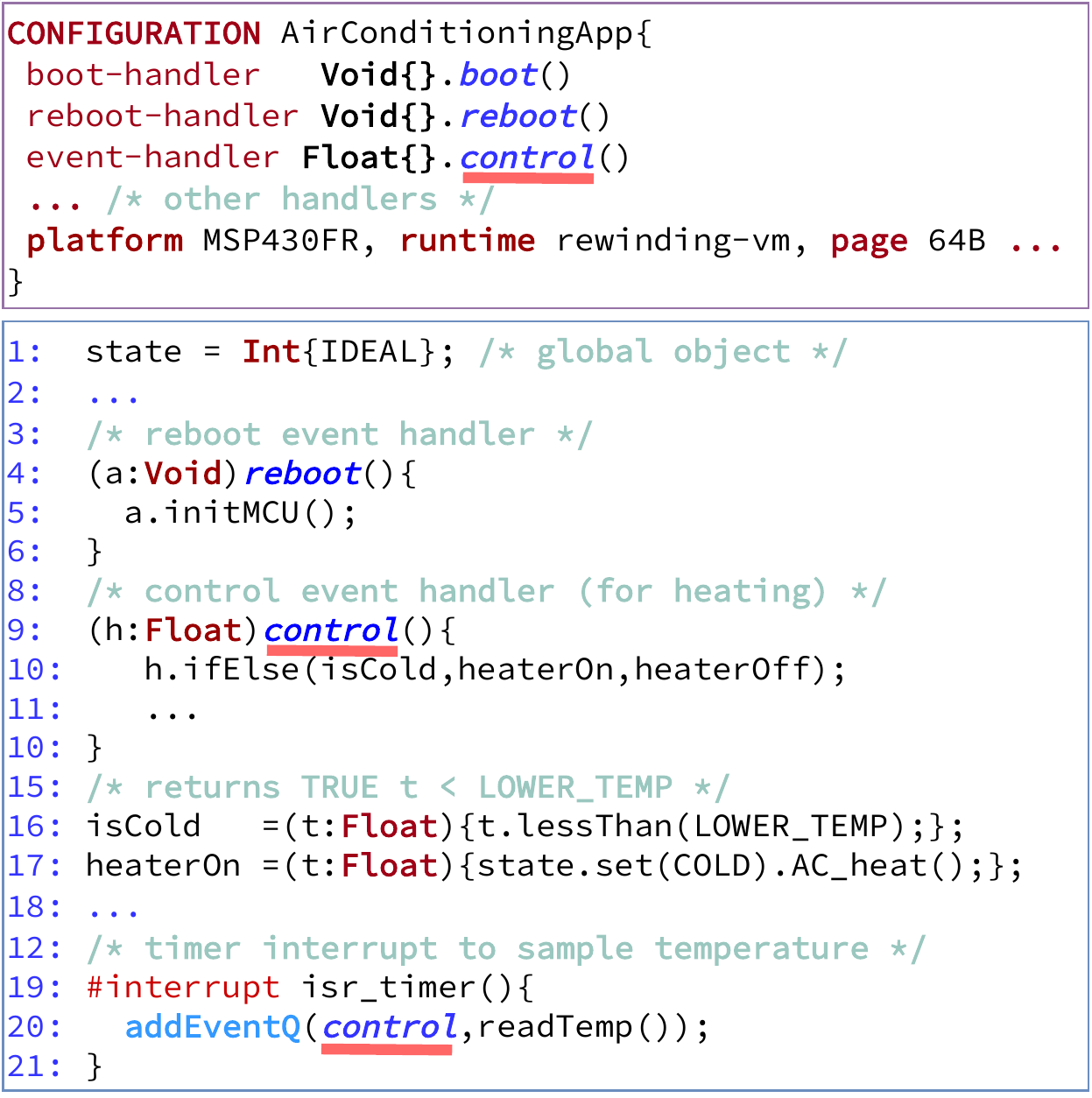} 
	\caption{Sample monitoring application in \langname. Application code contains all semantics of the computation, \vmname settings defining the events of the program and platform specific attributes. Interrupt code is for receiving the values from environment (e.g., sensors).
	}
	\label{fig:event_handling_configuration}
\end{figure}



\subsection{Putting Things Together: A Sample Sensing Application}

Figure~\ref{fig:event_handling_configuration} presents an event-driven air monitoring application, which includes an application source code and a configuration file. \langname compiler (implemented using Xtext~\cite{eysholdt2010xtext}) produces C code from the given \langname program. The generated code includes a single C source file and its header. The source file also contains the implementation of the target \vmname.  The \langname compiler requires a configuration file, which mainly contains the list of event handlers, the name of the target hardware platform, and some specific parameters of the selected \vmname implementation (such as non-volatile memory size, the size of the event queue). 

The application code contains the objects, methods, and interrupt handlers. The event handlers  \code{boot}, \code{reboot}, and \code{sleep} (which is not shown in the figure), are mandatory. The \code{boot} event occurs the first time the computer boots after being deployed. The \code{reboot} event occurs after recovery from a power failure, which triggers the reboot handler that restores the state of the computation. \vmname triggers the \code{sleep} handler when there is no event to be processed, which puts the processor in a low power mode to save energy. The timer interrupt handler (Lines 19--21) adds an event to the event queue of \vmname by calling \code{addEventQ} method with the sensed temperature value (via \code{readTemp}) and the corresponding event handler (which is \code{control} in this case) as parameters. \vmname processes this event by calling the \code{control} event handler (Lines 9--11), which processes the events generated from the timer interrupt service routine. Inside this routine, the heater is turned on or off based on the received temperature value (Line 10). 


\begin{figure*}
	\centering
	\includegraphics[width=\linewidth]{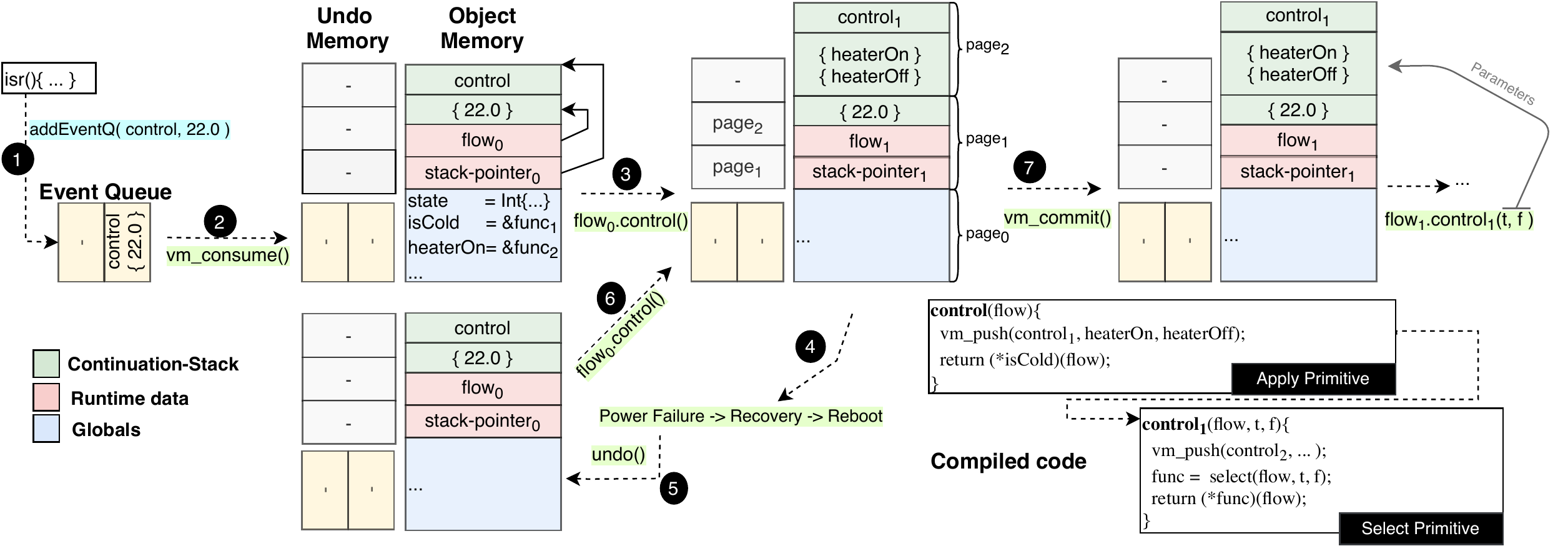} 
	\caption{The steps taken by \implI to execute  the \code{control} event described in Fig. \ref{fig:event_handling_configuration}. In the lower right corner of the figure, a part of the compiler generated code is presented. The compiler splits the \code{control} method into primitive functions and then recomposes into its basic blocks.}
	\label{fig:rewinding-in-action}
\end{figure*}

\section{\vmname Intermittent Virtual Machine}
\label{sec:vm}

\vmname is a single-input/single-output system that specifies a runtime environment for event-driven intermittent programming. \langname programs execute, without any modification, on different hardware and runtime environments conforming \vmname specification.

\vmname specification comprises an event queue, a non-volatile object memory, and a continuation executed by its runtime engine. \vmname pushes the events generated by the interrupt service routines to the event queue. \vmname removes the event at the head of the event queue and creates a continuation in object memory using that event.  As mentioned,  continuation represents the control state of the computer program, and it consists of a set of objects and methods to be applied. Running a continuation may create/return another continuation. \vmname runs the continuations until there is no returned continuation. When there is no event to consume in the queue, \vmname sleeps until an interrupt generates an event. 

\vmname state (which represents the computational state) is composed of the events in the queue, the continuation of the running event, and the global objects. The object memory region in non-volatile memory maintains the global objects and the running continuation.  Before calling the next function (i.e., running the subsequent continuation), \vmname can decide to persist the state in object memory to preserve forward computation and not to lose the intermediate results in case of power 
failures.

\vmname specifies the artifacts (event buffer, object memory, runtime engine, and running program) and their relationships abstractly. For example, the event buffer can be implemented as a regular first-in-first-out (FIFO) queue or a priority queue, as long as the system's single-input behavior is not violated. Different design choices can lead to different \vmname implementations. In the following section, we describe \implI, which is our main \vmname implementation. 

\subsection{\implI: An Undo-Logging \vmname}
\label{sec:vm1}

\implI stores the execution context in the continuation stack and keeps the object memory consistent across power failures via the undo-logging mechanism. Events in the event queue, global objects, and continuation stack in the object memory represent the computational state. The metadata about the continuation stack is stored in the \emph{runtime data} region of the object memory.

\noindpar{Event Handling.} \implI provides a queue for event buffering, which holds the event objects and the methods that need to be applied to these objects (\emph{Event Queue} in Fig.~\ref{fig:rewinding-in-action}). \implI runtime engine stores the continuations in a stack in object memory (\emph{Continuation Stack} in Fig.~\ref{fig:rewinding-in-action}). The runtime engine starts event execution by copying the event object (event-data) and event-method (event-handler) to the continuation stack. It is worth mentioning that \implI reduces the event handling to a single-producer-single-consumer problem which eliminates the race conditions on the event buffer. During execution, the runtime engine pops a method from the continuation stack and runs this method on the flow object. Control passes to the method, which can modify the objects. 

\noindpar{Undo-Log and Memory Consistency.} The undo-log mechanism is activated when modifying objects to preserve memory consistency. The modifications on the objects are done only by calling primitive functions. Every primitive function calls the runtime engine's log function before modifying an object. The object memory comprises blocks called pages. The programmer, for efficiency, may configure the page sizes of the object memory.  The log function copies the original page to undo-log memory before any modification on that page. When \implI reboots after a power failure, it copies all the logged pages (including the pages of runtime data) into their corresponding pages in the object memory and continues the program. This mechanism ensures the consistency of the object memory by eliminating partial modifications.  

\noindpar{Forward Progress and Execution Flow.} The method being executed can push other methods to the continuation stack. The method execution finally returns an object and gives the control back to the runtime engine. The runtime engine saves the returned object as a flow object. The runtime atomically clears the undo-log memory as the last operation. In this way, the runtime keeps memory consistent and guarantees the forward progress of the program. 

\noindpar{I/O Handling.} The \langname compiler already splits the program code blocks with I/O primitive functions into three sections, as described in Section \ref{sec:io:primitive}. This strategy already ensures the atomic execution of I/O operations. Therefore, the \implI runtime engine does not treat I/O operations in a specific way.  

\noindpar{\implI Compiler Optimizations.} We implemented two compiler optimizations for \implI: (i) executing several primitive functions as a block, and (ii) some loop optimizations to reduce \vmname overheads such as repetitive undo logging (i.e., page copy) operations. Programmers can modify \vmname application configuration file to enable optimizations. 

\subsubsection{An Example \implI Execution.} 
Fig. \ref{fig:rewinding-in-action} shows how \implI handles the \code{control} event described in Fig. \ref{fig:event_handling_configuration}. In the first step, the interrupt service routine adds the address of the control method to the event queue along with the sensed temperature value (which is a floating-point value of 22.0). In the second step, the VM copies the event from the event queue to the object memory (depicted as \code{control} and \code{22.0} in the continuation stack). In the lower right corner of Fig. \ref{fig:rewinding-in-action}, a part of the compiler-generated code is presented. The compiler splits the \code{control} method into primitive functions and then recomposes into its basic blocks.  As can be seen from the figure, the recomposed version of the \code{ifElse} method (Line 10 in Fig. \ref{fig:event_handling_configuration}) is fragmented into two continuations that represent \code{apply} and \code{select} primitives. In the third step, the VM removes the first method from the continuation stack  (\code{control} function in the lower right corner of Fig. \ref{fig:rewinding-in-action}) and runs it.  Since this process changes the runtime data and the stack of the non-volatile memory, it copies the pages to undo memory before any update. The fourth step in the figure shows that the computer restarts after a power failure. In the fifth step, the VM calls the undo function because it detects that the undo memory is not empty. It brings the non-volatile object memory to the last consistent state. Then, in step 6, the method on the stack runs, as described in step three. In the seventh step, the undo memory is cleared by committing it. Forward progress of the computation is ensured by executing the operations  \code{vm\_consume}, \code{vm\_commit} and \code{undo} atomically. In the next steps (not shown in the figure), the remaining methods in the continuation stack execute. They pass the returned flow objects to the next methods till the continuation stack becomes empty.

\subsection{Other \vmname Implementations}

We also implemented two different \vmname versions, named \implII and \implIII. \implII does not contain an undo log memory and requires hardware support to capture an interrupt when the voltage value on the capacitor crosses the lower-threshold voltage. This interrupt persists the computational state in non-volatile memory and puts the runtime into sleep mode to stop computation. This strategy prevents memory inconsistencies without the need for an undo-log. \implII's overhead is lower than \implI's since \implII does not include page copy operations between log memory and object memory. We implemented \implIII to test any \langname program on a personal computer with continuous power. This implementation allows us to test the correctness of the program logic without loading the code on a microcontroller. 


\section{Evaluation}
\label{sec:eval}

We evaluated our \vmname implementations considering three compute-intensive applications Cuckoo Filter (CF), Activity Recognition (AR), and  Bit Count (BC). These applications are used as the de facto benchmarks by most of the earlier studies on intermittently powered devices (\cite{kortbeek2020TICS,alpaca,colin2016chain,yildirim2018ink}). CF stores and reads an input data stream using a cuckoo filter with 128 entries and searches the inserted values. AR classifies accelerometer data, computing the mean and standard deviation of a window of accelerometer readings to train a nearest-neighbor model to detect different movements. We used a window size of three and read 128 samples from each class (shaking or stationary) in the training phase. BC counts the number of 1s in a bitstream with seven different methods that are executed 100 times each.

We compiled these applications using the \implI and \implII compilers and executed them on the MSP430FR5994 evaluation board \cite{msp430fr5904}. To power MSP430 evaluation boards intermittently, we used Powercast TX91501-3W transmitter \cite{powercast}, which emits radio frequency (RF) signals at 915 MHz center frequency. P2110- EVB receiver, connected to our MSP430FR5994 evaluation board, harvests energy from these RF signals. We placed the P2110- EVB receiver at a distance of 60cm from the RF transmitter.

\subsection{Execution Time Overhead}
We evaluated the performance of \vmname implementations of the benchmarks on harvested energy and continuous power by considering their execution times. For comparison, we used their InK \cite{yildirim2018ink} implementations since InK is one of the de facto task-based systems for intermittent computing. We directly used InK-based implementations of the benchmarks (CF, AR, and BC) from the InK repository\cite{tudssl_2019}. Since the tasks have fixed sizes in InK, they may exceed the energy buffer of the device, or they may perform inefficiently due to frequent task transitions (causing redo logging of all shared variables to preserve memory consistency). To re-calibrate the size of the tasks, the programmer must recompose all tasks for the new device. This way of programming limits the code portability.

\begin{figure}
	\centering
	\includegraphics[width=1\linewidth]{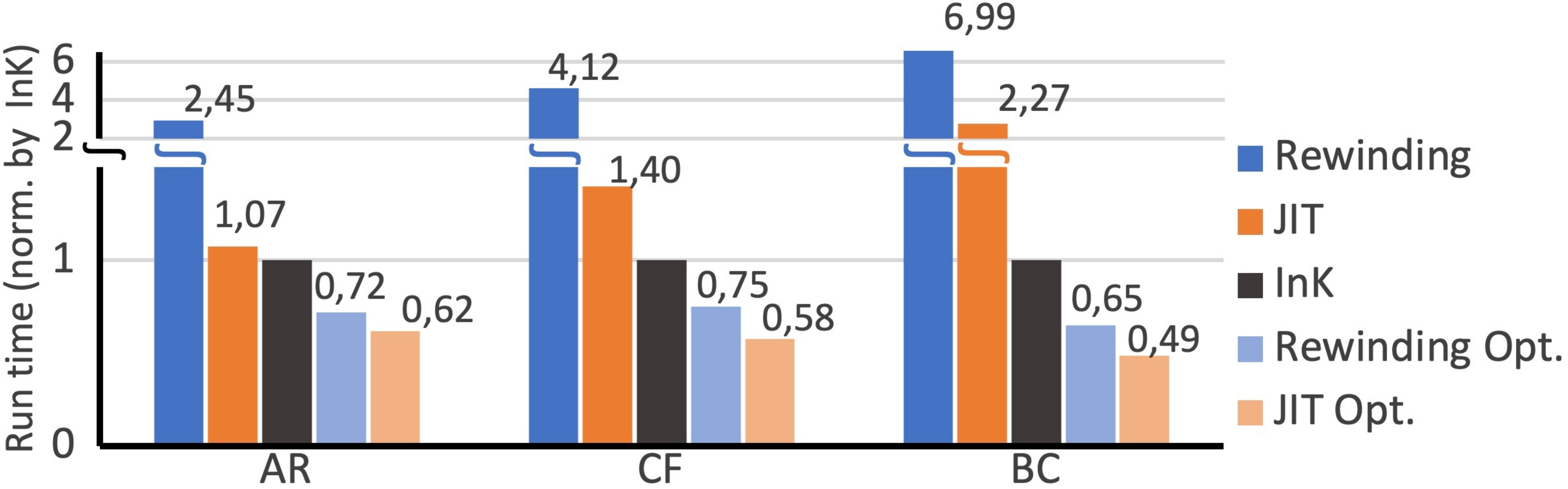} 
	\caption{Normalized execution times of AR, CF and BC benchmarks with InK, \implI and \implII under continuous power.}
	\label{fig:execution_time_1}
\end{figure}

\begin{figure}
	\centering
	\includegraphics[width=1\linewidth]{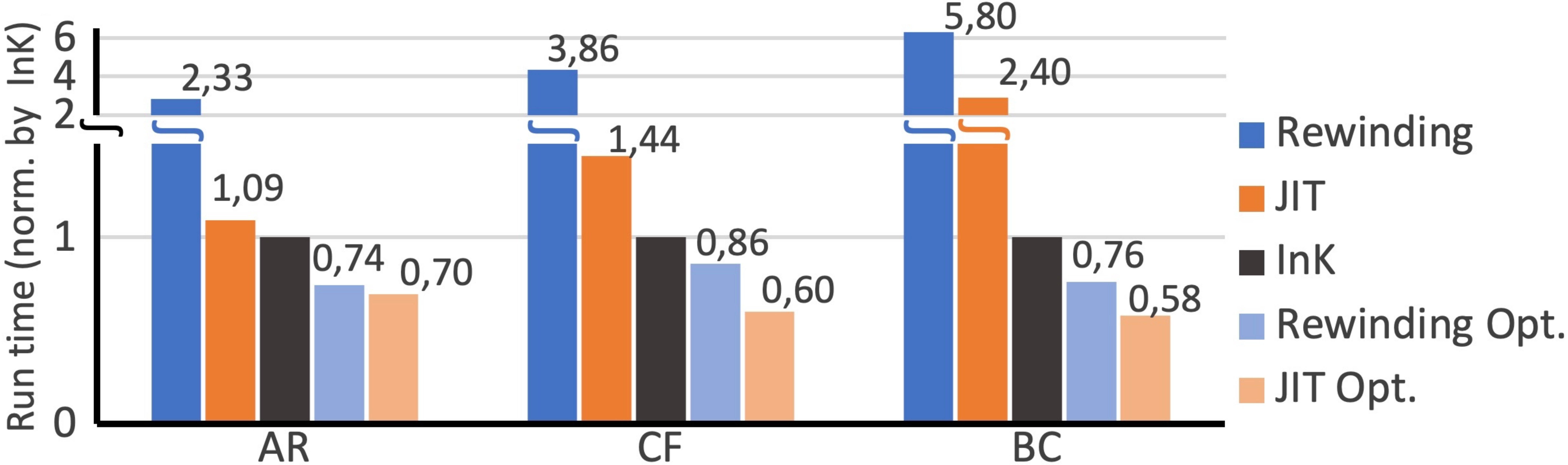} 
	\caption{Normalized execution times of AR, CF and BC benchmarks with InK, \implI and \implII under RF energy harvesting scenario.}
	\label{fig:execution_time_2}
\end{figure}

Fig.~\ref{fig:execution_time_1} and Fig.~\ref{fig:execution_time_2} show the normalized execution times of the benchmarks with InK and \vmname (\implI, \implI optimized, \implII, \implII optimized) on continuous power and intermittent execution on RF-power. The compiled code of \implI and \implII can run on the devices with a minimal energy buffer because the compiler recomposes primitive functions regarding basic blocks. Recomposing the code into its basic blocks leads to small continuations and consequently more (continuation) stack operations in \implI and \implII, and more undo logging operations in \implI. 
Optimized versions of applications are composed of continuations with more operations. Therefore, these applications run more efficiently compared to InK apps. We applied the loop optimizations to all loops and branch optimizations to some branch method invocations in these applications. 

\begin{figure}
	\centering
	\includegraphics[width=1\linewidth]{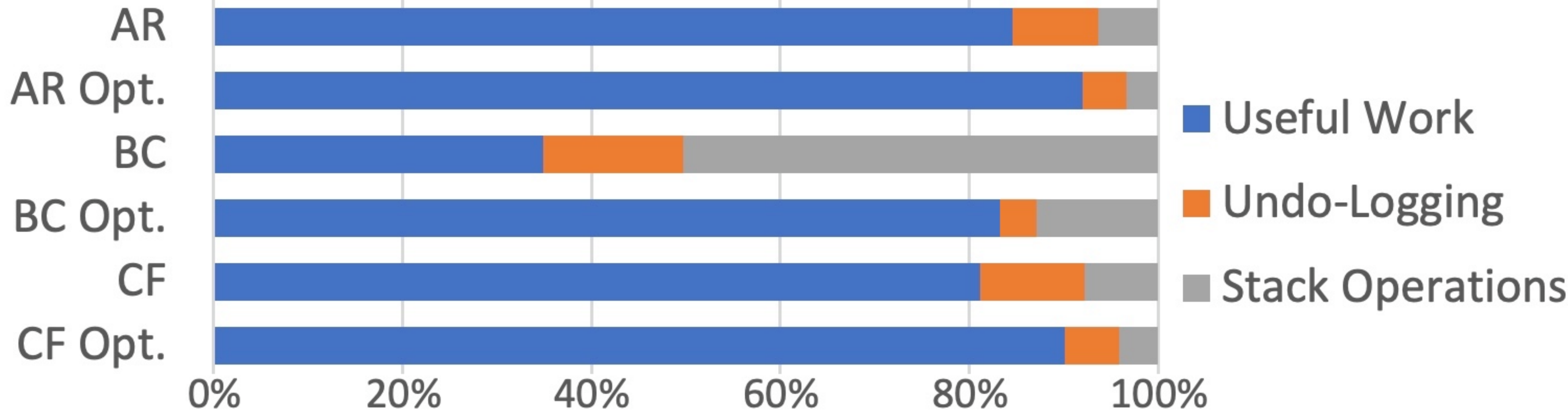} 
	\caption{RewindingVM overhead, split per operation.}
	\label{fig:rew_portions}
\end{figure}

\begin{figure}
	\centering
	\includegraphics[width=1\linewidth]{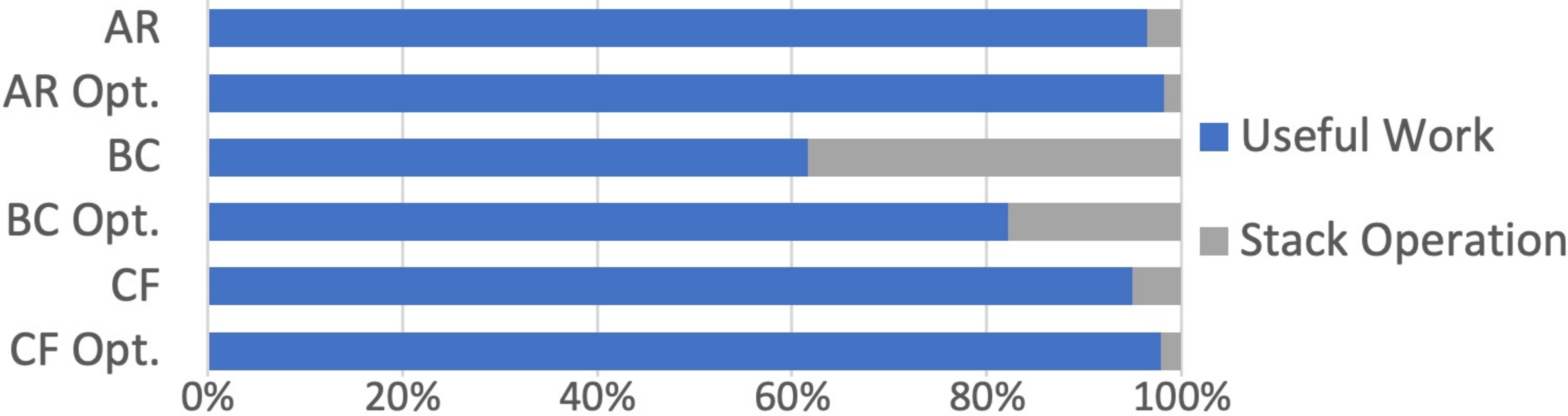} 
	\caption{JustInTimeVM overhead, split per operation.}
	\label{fig:jit_portions}
\end{figure}

\subsection{\vmname Point-to-Point overheads}

Fig.~\ref{fig:rew_portions} and Fig~\ref{fig:jit_portions} show the useful work and overheads (i.e., undo-logging and stack operations) of corresponding runtimes on continuous power. \implI and \implII uses a stack to run the continuations. When there is more branching in the program, the continuation stack operation creates more overhead because \implI and \implII run the basic blocks in one cycle, which means every branch needs continuation stack operation. In BC, the overhead of stack manipulation is higher due to many branch operations, which also causes the worst performance compared to other benchmarks. On the other hand, loop optimizations in BC had the most impact on the performance compared to other benchmarks, increasing the performance by a factor of 10.


Before modifying a page that has not been logged before, the undo-logging mechanism is triggered, which introduces page search and page copy overheads. Since the log memory size is small for the benchmarks, we chose a sequential page search in the log memory. Apparently, the page size affects the undo logging performance. The virtual machine configuration file of \implI and \implII contains a page size setting. The page size of 32 and 64 bytes gave the best performance in these applications.

\begin{table}
\centering
\caption{Memory consumption for three benchmark applications written in InK and \vmname.}
	\label{table:memory_comparison}
\begin{tabular}{ccccccc}
\toprule
  \textbf{App.}& \textbf{Memory(B)}& \textbf{InK} & \textbf{Rew.} & \textbf{Rew.Opt.} & \textbf{JIT} & \textbf{JIT.Opt.}\\
\toprule
\multirow{2}{1em}{AR} & .text & 3822 & 8624  & 6810  & 7950  & 6136\\
                      & .data & 4980 & 694   & 694   &  420  & 420\\
\multirow{2}{1em}{BC} & .text & 3518 & 11258 & 9928  & 10714 & 9378\\
                      & .data & 4980 & 882   & 882   & 676   & 548\\
\multirow{2}{1em}{CF} & .text & 2708 & 11980 & 10014 & 11302 & 9350\\
                      & .data & 5213 & 886   & 694   & 804   & 420\\
\bottomrule
\end{tabular}
\end{table}

\subsection{\vmname Memory overheads}
Table \ref{table:memory_comparison} shows the memory overheads of InK and \vmname implementations. Since primitive functions are translated into C codes, PureVM programs have larger code sizes than InK programs. InK uses global shared variables for communication among functions, and hence it has larger data memory than \vmname. The programmer configures data memory of \implI and \implII via the configuration file. The stack size required by virtual machines contributes to their data memory requirements.  The code size increase is the cost which \vmname implementations pay for their performance and platform independence. However, overall results show that the memory requirement of \vmname is comparable with InK's memory requirement.

\begin{figure}
	\centering
	\includegraphics[width=0.6\columnwidth]{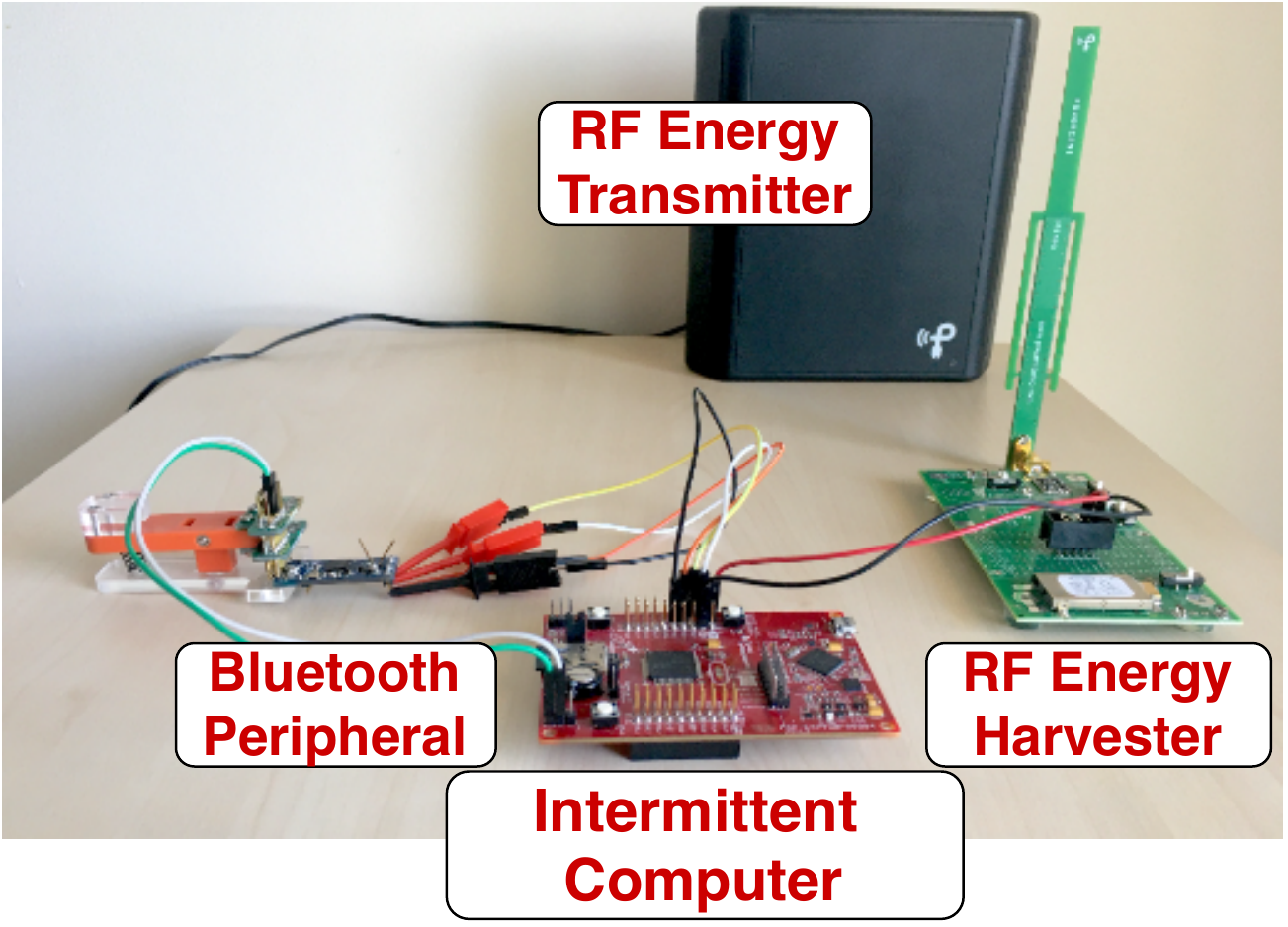} 
	\caption{Our testbed deployment used to evaluate our HVAC application.}
	\label{fig:AirConditioner}
\end{figure}

\subsection{Case Study: Heating, ventilation, and air conditioning  (HVAC) controller}
As a case study to demonstrate the applicability of \vmname, we developed an air condition controller for home automation (Figure ~\ref{fig:AirConditioner}). The goal is to get the room temperature frequently enough and send a message to home automation to keep the room temperature in the ideal temperature range. The application uses the \vmname event mechanism to reduce energy consumption as much as possible: after controlling the room temperature, the application switches to sleep mode. A reboot from a power failure or a timer interrupt (with 30-second intervals) triggers the program, which estimates the room temperature using an analog-digital converter configured to the internal temperature sensor of the microcontroller. If the room temperature is not ideal, the application starts asynchronous communication via Nordic nRF52832 \cite{nRF52832} (which supports BLE) using interrupts over the serial peripheral interface. We ran this application for 3 hours. We observed that the temperature of the environment was measured 294 times, and 22 BLE advertisement messages were sent to the HVAC system. During the entire run, there were 18 power failures and recovery.
\section{Conclusion and Future Work}
\label{sec:conc}

In this work, we introduced a new virtual machine (\vmname) that abstracts a transiently powered computer and a new continuation-passing-style programming language (\langname) used to develop programs that run on \vmname. This two-layer structure provided a loosely coupled architecture that facilitates the development of platform-independent and reusable event-driven sensor applications. We believe that this is a significant attempt for virtualizing intermittent computing.

As follow-up work, we plan to add new language constructs to \langname to handle the expiration of sensor readings. Due to long charging times after power failures, sensed data might lose its validity. In this case, the sensor value becomes useless and can be discarded. While such a requirement is absent in continuous computing, it exists in intermittent computing~\cite{kortbeek2020TICS,hester2017mayfly}. We also plan to port \vmname to different ultra-low-power micro-controllers and introduce more sophisticated compiler optimizations. As of now, \vmname does not implement any task scheduling mechanism. We leave integrating scheduling mechanisms, e.g., real-time scheduling of tasks ~\cite{karimi2021real,maeng2020CatNap}, to \vmname as future work.

\bibliographystyle{IEEEtran}
\bibliography{references}

\end{document}